\documentstyle[preprint,aps]{revtex}
\begin{document}
\title{\bf Duality in two capacitively coupled layered arrays of
                    ultrasmall Josephson junctions}
\author{Jorge V. Jos\'e}
\address{
Physics Department and Center for the Interdisciplinary
Research on Complex Systems \\ Northeastern University, Boston, MA,
02115, USA}
\maketitle
\begin{abstract}
We consider the problem of two capacitively coupled 
Josephson junction arrays  made of  ultrasmall junctions.
Each one of the arrays can be in the semiclassical or quantum
regimes, depending on their physical parameter values.
The former case is dominated by a Cooper-pair superfluid 
while the quantum one is dominated by  dynamic vortices leading to 
an insulating behavior. We first consider the limit when both arrays are in the
semiclassical limit, and next  the case when one array is
quantum and the other semiclassical. We present  WKB and Mean Field theory 
results for the critical temperature of each array when  both  are 
in the  semiclassical limit. When one array is in the semiclassical regime and 
the other one in the quantum fluctuations dominated regimes, we 
derive a duality transformation between the charged and
vortex dominated arrays that involve a gauge vector field,
which is proportional to the site coupling capacitance
between the arrays. The system considered
here has been fabricated and we make some predictions as to possible
experimentally  measurable quantities that could be compared with
theory.
\end{abstract}

\newpage

\vskip 0.4cm
\centerline{\bf Personal note on Leo Kadanoff}
\vskip 0.4cm

I am delighted to be able to add this contribution  in honor of Leo 
Kadanoff's sixtieth birthday. I have purposefully chosen a subject that involves {\it duality 
transformations}.
Although in a completely different context, this goes back to several 
ideas I first learned from Leo many years ago, when I started my research career
under his tutelage. Duality transformations  continue  
to lead to new and  interesting physics. Leo has been involved with 
{\it duality transformations} in several  different contexts, making 
seminal contributions  to their understanding and applications.

I spent about three years with Leo, two as a his 
Ph.D. graduate student and one as 
his postdoctoral fellow. During this time I was able to see a real Master 
at work. Although we did not meet very often, every time we did (and he
was always available to meet me),  his insights and quickness left a 
deep impression on me. Everybody has his/her own style of doing research,
and while I could not claim to do research as Leo does, his approach, style 
and his depth made a lasting impression on my way of looking at physics 
and physics problems addressing their detailed quantitative solutions.

\section{\bf Introduction}
\label{sec:intro}

The subject treated in this paper relates to interesting quantum properties of
Josephson junction arrays (JJA) made with ultrasmall junctions. 
Layered Josephson junction arrays  have been 
the source of many theoretical and experimental studies in the last 
few  years \cite{rev}. Recent advances in submicrometer technology 
have made it possible to fabricate relatively large arrays of ultrasmall 
superconductor-insulating-superconductor (SIS) 
Josephson junctions \cite{herre2,tighe,chalmers,3degas}. The areas of 
these junctions can vary from a few microns to submicron sizes. 
Under these circumstances the long range phase coherent properties 
of the JJA depend crucially on the interplay between  the Josephson energy, 
$E_J$, and charging energy, $E_C$. 
Detailed experiments have been carried out, for example at Delft,
 that have produced a phase diagram of temperature vs the quantum parameter 
${\alpha \equiv \frac{E_C}{E_J}}$ \cite{herre2}.  We have calculated the 
$\alpha$ vs temperature phase diagram and we made \cite{rojas_prb,rojas}
 a direct successful 
comparisons to the experimental results \cite{herre2}. Our results were 
obtained using  a WKB-renormalization-group approach, plus a variational
and quantum Monte Carlo (QMC) calculations \cite{rojas_prb}. An important QMC 
result is that there appears to be a low temperature QUantum fluctuation 
Induced Transition  ($QUIT$) in this system  \cite{jacobs,rojas_prb}.

For the most part the experimental systems have been two-dimensional, 
but prototype quasi-three-dimensional 
samples have also been fabricated \cite{sohn}. In this paper quasi- 
means two layers of JJA capacitively coupled at each lattice site. 
There are two dominant  contributions to the charging energy in the type of 
junctions fabricated \cite{sohn}; one due to the addition of a 
 Cooper pair charge to a superconducting island given by 
$E_{C_{\rm s}}=\frac{2e^2}{C_{\rm s}}$ with $C_s$ the self capacitance
and $e$ the electronic charge,
and the charging  energy necessary to transfer a Cooper charge from one
island to its  nearest neighbor, given by 
$E_{C_{\rm m}}=\frac{2e^2}{C_{\rm m}}$, with $C_m$ the mutual capacitance.
In the Delft experiments,  $C_{s}\sim 3\times 10^{-17}$F, 
and $C_{m}\sim 1\times 10^{-15}$F, which means that
$C_m$ can be two orders of magnitude larger than $C_s$.

The phase diagram for one layered JJA has the following 
general characteristics. At low temperatures, for small 
$\alpha$ there is a superconducting
phase in which the Cooper pair charges are delocalized, while
for large $\alpha$  the system has delocalized vortices and the array
is an insulator. There is a phase boundary that separates the 
superconducting to insulating  regions. Now assume that we have two
layered JJA capacitively coupled at each lattice site. This configuration is 
potentially quite interesting  since, as mentioned above, each array 
can be in one of two extreme limits; one Cooper charge dominated and 
the other vortex dominated.  Each array is now described by
its quantum ratio  $\alpha _i=\frac{E_{C_i}}{E_{J_i}}$ ($ i=1,2$).
We can then imagine to have the two arrays in four
possible configurations.
When $\alpha _i\ll 1$,  the i-th array is dominated by localized
vortex excitations, $V_i$, while the Copper pair excess charge
excitations, $Q_i$, are in a superfluid or superconducting
state. In the $\alpha _i\gg 1$ regime the array has the $Q_i$'s 
localized in an insulating state while 
the $V_i$'s are delocalized. We can also have the extreme cases when
both arrays are semiclassical or in the quantum regimes.

In this paper we shall consider first the case where both arrays are in the
semiclassical regime. We obtain an effective partition function that
allows to calculate the change of each array critical temperatures
as a function. This type of analysis was quite informative in the 
one-array problem. Next we move to consider the general case where
we derive a Hamiltonian that is valid for all parameter regimes. This
Hamiltonian is quite complex and difficult to analyze in full. Instead
we consider the interesting case when one of the arrays is charge and the 
other vortex dominated. After a series of transformation we arrive at
an effective Hamiltonian that exhibits interesting duality properties.
We analyze the imaginary time dynamics of this Hamiltonian in the case
where only one vortex in one array and one charge in the other are considered.
Here we show that these tow excitations interact via a gauge-like interaction
proportional to the interaction capacitance between the arrays.

The outline of the paper is the following. in Section III the case where 
both arrays are in the semiclassical limit since in that case we can 
analyze in some detail the changes in the critical temperature of both 
arrays independently via a variational Mean Field analysis. In Section 
IV  we consider the interesting limit when
one array is semiclassical and the other quantum. In this case we can
derive an effective action for the problem that allows a general
analysis of the interaction of a vortex in one array with a charge in
the other via an interaction term that has the form of a minimal gauge
coupling proportional to the interaction capacitance. 
The other extreme case where both arrays are quantum is not considered 
here since that case is harder to analyze. We conclude the
paper with some conclusions in Section V.

In this paper I present some results of work done in collaboration with 
C. Rojas (related details can be found in  \cite{rojas_prb} and in his 
Ph.D. thesis \cite{rojas}).

\section{The Model}

\label{section:two-layers}
%

%
%
%
In this section we  define the model that describes 
 a quasi-three-dimensional array composed of two JJA layers coupled
at each site by an ultrasmall  capacitor. In our analysis we will have in mind
the prototypical  samples fabricated at Delft \cite{sohn}.   
In these samples the size for each layer were $L_x=230$, and $L_y=60$.
The typical parameter for the intra-layer mutual
capacitance was $C_m\approx 2.3 fF$, and the interplane local 
interaction capacitance was $C_{\rm int} \approx 0.6 fF$.
%
%
The model consists of two planar arrays stacked on top of each other. The 
intra-array interaction between the superconducting islands in each array
contains  an electrostatic and a Josephson coupling. The two planes are 
only capacitively coupled. The model Hamiltonian is given by
\begin{equation}
\label{initial-hamiltonian}
    \hat{H} = \frac{q}{2}\sum_{\vec r}\sum_{\mu} \hat{\bf V}_\mu(\vec r)
              \hat{n}_\mu(\vec r) + F_1(\{\hat\phi_1\})+F_2(\{\hat\phi_2\}),
\end{equation}
where the index $\mu=1,2$ 
labels the two arrays. The operators $\hat {\phi}_\mu$ and ${\hat n}_\mu$
satisfy the commutation relations
$
[\hat n_{\mu}(\vec r_1),\hat\phi_{\mu '}(\vec r_2)]
=-i\delta_{\vec r_1,\vec r_2}\delta{\mu,\mu '}.
$
The functions $F_\mu$ are the Josephson 
interaction terms,
\begin{equation}
\label{def-F's}
        F_\mu(\{\hat\phi\}) = E_J^{(\mu)} \sum_{<\vec r_1,\vec r_2>}
                   (1-\cos(\hat\phi_\mu(\vec r_1)-\hat\phi_\mu(\vec r_2))).
\end{equation}
Here $E_J^{(\mu)}$ is the Josephson energy coupling constant, for the 
junctions in array $\mu$. ${\bf V}_\mu(\vec r)$ is the electrostatic 
potential felt by the charges contained in the superconducting island 
located at $\vec r$ in array $\mu$. This potential is produced by 
all the other charges in both arrays, and it is obtained  from the  
discrete Poisson equation
\begin{equation}
\label{equation-for-v}
    q\ n_1(\vec r) = C_{\rm m}^{(1)} \sum_{\vec u} \left[{\bf V}_1(\vec r)
                     - {\bf V}_1(\vec r+\vec u)\right] + C_{\rm s}^{(1)} 
                     {\bf V}_1(\vec r) + C_{\rm int} \left[{\bf V}_2
                     (\vec r)-{\bf V}_1(\vec r)\right].
\end{equation}
Here the $\vec u$ summation is over nearest neighbors.
In a uniform square lattice $\vec u = \{\pm\hat x,\pm\hat y\}$.
The complementary equation for $n_2(\vec r)$ is obtained
by interchanging ($1\leftrightarrow 2$). In Eq.(\ref{equation-for-v}),  
$C_{\rm s}^{(\mu)}$ is the self capacitance of a superconducting island 
in array $\mu$. The previous equation can be written in the compact form,
\begin{equation}
\label{short-eq-for-v}
       q\ n_\mu(\vec r_1) = \sum_{\nu} \sum_{\vec r_2} {\bf C}_{\mu,\nu}
                            (\vec r_1,\vec r_2) {\bf V}_\nu(\vec r_2),
\end{equation}
where the capacitance supermatrix ${\bf C}_{\mu,\nu}$ is  made of four 
blocks labeled by $\mu,\nu=1,2$,
\begin{equation}
\label{block-matrix}
     {\bf C}_{\mu,\nu}(\vec r_1,\vec r_2) = 
                \left\{ \begin{array}{ll}
                        \left(C_{\rm s}^{(\mu)}+zC_{\rm m}^{(\mu)}+C_{\rm int}
                        \right),
                                    & \mbox{if $\mu=\nu$ and $\vec r_1=
                                                              \vec r_2$},\\
                        -C_{\rm m}^{(\mu)}, 
                                    & \mbox{if $\mu=\nu$ and $\vec r_1=
                                                        \vec r_2\pm\vec u$},\\
                        -C_{\rm int}, 
                                    & \mbox{if $\mu\ne\nu$ and $\vec r_1=
                                                                \vec r_2$},\\
                          0,
                                    & \mbox{otherwise.}
                        \end{array}
                \right.
\end{equation}      
The diagonal blocks of this matrix are the intra-array capacitance 
matrices. We will use the following notation for them
\begin{equation}
\label{intra-matrix}
      {\bf C}_{\mu} = {\bf C}_{\mu,\mu}.
\end{equation}
The off-diagonal parts of the ${\bf C}_{\mu,\nu}$ supermatrix 
are given by block matrices proportional to the identity matrix 
$-C_{\rm int} {\bf I}_{N,N}$, with $N$ the linear size of the arrays. 
The inverse matrix of 
Eq.(\ref{block-matrix}), ${\bf\tilde C}$, is obtained from solving the equation
\begin{equation}
\label{def-of-c-tilde}
    \sum_{\nu} \sum_{\vec r_2} {\bf\tilde C}_{\mu,\nu}(\vec r_1,\vec r_2)
               {\bf C}_{\nu,\rho}(\vec r_2,\vec r_3) =
                \delta_{\mu,\rho}\ \delta_{\vec r_1,\vec r_3}.
\end{equation}
The explicit components of the inverse matrix are  given by
\begin{eqnarray}
\label{c-tilde-11}
    & &{\bf\tilde C}_{1,1} = {\bf C}_1^{-1}\left[{\bf I} - C_{\rm int}^2
       {\bf C}_2^{-1}{\bf C}_1^{-1}\right]^{-1},\\               
\label{c-tilde-22}
    & &{\bf\tilde C}_{2,2} = {\bf C}_2^{-1}\left[{\bf I} - C_{\rm int}^2
       {\bf C}_1^{-1}{\bf C}_2^{-1}\right]^{-1},\\
\label{c-tilde-12}
    & &{\bf\tilde C}_{1,2} = C_{\rm int} {\bf C}_1^{-1} {\bf C}_2^{-1}
       \left[{\bf I}-C_{\rm int}^2{\bf C}_1^{-1}{\bf C}_2^{-1}\right]^{-1},\\
\label{c-tilde-21}
    & &{\bf\tilde C}_{2,1} = C_{\rm int} {\bf C}_2^{-1} {\bf C}_1^{-1}
       \left[{\bf I}-C_{\rm int}^2{\bf C}_2^{-1}{\bf C}_1^{-1}\right]^{-1}.
\end{eqnarray}
For a uniform array an explicit expression for this matrix can be found 
using a Fourier representation of the matrices.

Finally, using Eqs. (\ref{initial-hamiltonian}), (\ref{short-eq-for-v}), and 
(\ref{def-of-c-tilde}), we can write down the model Hamiltonian studied in this
paper as
\begin{equation}
\label{hamiltonian-with-c-tilde}
     \hat{H} = \frac{q^2}{2} \sum_{\vec r_1,\vec r_2} \sum_{\mu,\nu}
     \hat{n}_{\mu}(\vec r_1) {\bf\tilde C}_{\mu,\nu}(\vec r_1,\vec r_2)
     \hat{n}_{\nu}(\vec r_2) + F_1(\{\hat\phi_1\}) + F_2(\{\hat\phi_2\}).
\end{equation}
\section{Semiclassical capacitively coupled arrays}
\label{sec:semiclassical}
In this section we begin our analysis of the quasi-three dimensional
JJA model 
defined by Eq.(\ref{hamiltonian-with-c-tilde}), when
the parameters in both arrays are in the small $\alpha$ 
semiclassical regime. Here we
estimate the change in the critical temperature, $T_c$, for each array
as a function of  $C_{\rm int}$. We use a WKB semiclassical expansion 
valid for small  $\alpha$ values, as we did in the single 
array problem  \cite{jacobs,rojas_prb}. In this limit we first
obtain an effective action that we then analyze within a variational
Mean Field theory (MFT) approach. 
In the prototypical  Delft samples,  they had  small ratio values for
$(C_{\rm int}/C_{\rm s})$. To have a consistent
semiclassical expansion in $C_{\rm int}$, as follows from
looking at Eqs.(\ref{c-tilde-11})-(\ref{c-tilde-21}), 
we need to carry out the expansion at least to second 
order in ${\bf\tilde C}$, which is equivalent
to doing  the expansion up to second order in  $q^2$. After a long, 
but direct, calculation we obtain the effective semiclassical partition 
function,
\begin{equation}
\label{two-layer-partition-function}
      Z_{SC} = \int d\overline{\phi}\ \exp\{-S_{\rm eff}[\overline{\phi}]\},
\end{equation}
with the effective semiclassical action up to second order given by
\begin{eqnarray}
\label{eff-action-two-layers}
      S_{\rm eff}[\phi] &=& \beta \sum_{\mu} \Bigg\{1-\frac{\beta q^2} 
      {12}\left[{\bf\tilde C}_{\mu,\mu}(|\vec 0|)-{\bf\tilde C}_{\mu,\mu}
                (|\vec u|)\right] +\nonumber\\
       & &\ \ \ \ \ \ \ \ \ 
          + \frac{(q^2\beta)^2}{1152}
          \left[{\bf\tilde C}_{\mu,\mu}(|\vec 0|)-{\bf\tilde C}_{\mu,\mu}
                (|\vec u|)\right]^2\Bigg\}\ F_\mu(\phi_\mu) \nonumber\\
       & & + \frac{(q^2\beta)^2\beta^2}{720} {\bf Tr}\left\{
           \sum_{\mu,\nu} {\bf\tilde C}_{\mu,\nu}\ \frac{\partial^2 F_\nu}
           {\partial\phi_\nu^2}\ {\bf\tilde C}_{\nu,\mu}\ \frac{\partial^2
            F_\mu}{\partial\phi_\mu^2}\right\}.
\end{eqnarray}
As a check, we note  that we recover the one-layer results if we  
keep terms up  to first order in $q^2$. We find that the superconducting 
state is stable at low temperatures, because the second-order 
contribution in $q^2$ has a negative  sign compared to the first-order 
contribution to the effective action. Notice that higher orders in  
$q^2$ coincide with higher orders in $\beta$. This 
conclusion does not, by itself, eliminate the possibility of having a 
$QUIT$ in the two layered problem.

Eq.(\ref{eff-action-two-layers}) does not have a simple form as in
the one JJA layer first-order expansion in $q^2$. The third term in 
Eq.(\ref{eff-action-two-layers}) has nonlocal interactions, 
which makes a direct calculation of the critical temperature 
in general complicated. To estimate the change 
in the critical temperatures as a
function of  $C_{\rm int}$, we have performed a couple of distinct MFT 
variational calculations for the partition function  given in 
Eq.(\ref{eff-action-two-layers}). 

For the general variational calculation first we split the effective  
action into two parts
\begin{equation}
\label{spliting-of-s-eff}
   S_{\rm eff}[\phi_1,\phi_2] = S_0^{(1)}[\phi_1] +
   S_0^{(2)}[\phi_2] + \left( S_{\rm eff}[\phi_1,\phi_2]
   - S_0^{(1)}[\phi_1] - S_0^{(2)}[\phi_2]\right),
\end{equation}
which gives the exact semiclassical expansion for the partition function 
\begin{eqnarray}
\label{exact-z}
    Z &=& Z_0^{(1)}\ Z_0^{(2)}\left<\exp\left\{ -S_{\rm eff}[\phi_1,\phi_2] 
        + S_0^{(1)}[\phi_1] + S_0^{(2)}[\phi_2]\right\}\right>_0,\\
\label{def-of-z0}
   Z_0^{(\mu)} &=& \int d\phi_\mu\ \exp\left\{-S_0[\phi_\mu]\right\}.
\end{eqnarray}
The average $< >_0$ is defined as
\begin{equation}
\label{def-of-<>0}
   \left<A\right>_0 = \frac{1}{Z_0^{(1)}Z_0^{(2)}} \int d\phi_1\ d\phi_2\
           A(\phi_1,\phi_2)\ \exp\left\{-S_0^{(1)}[\phi_1]-S_0^{(2)}
           [\phi_2]\right\}.
\end{equation}
We now use the variational inequality 
$\left<\exp\{A\}\right>\le\exp\{\left<A\right>\}$  to write
\begin{equation}
\label{approx-z}
   Z \le Z_0^{(1)}\ Z_0^{(2)}\ \exp\left\{\left<-S_{\rm eff}[\phi_1,\phi_2] 
           + S_0^{(1)}[\phi_1] + S_0^{(2)}[\phi_2]\right>_0 \right\}.
\end{equation}
The corresponding  variational free energy is then
\begin{equation}
\label{def-of-F-var}
   \beta F_{\rm var} = -\ln Z_0^{(1)}-\ln Z_0^{(2)}-\left<-S_{\rm eff}
                         [\phi_1,\phi_2] + S_0^{(1)}[\phi_1] + 
                         S_0^{(2)}[\phi_2]\right>_0.
\end{equation}
We next need to specify the functions $S_0^{(\mu)}$. They can be 
any general functions, but here we  restrict them to be 
functions of only one phase variable for each layer. These functions 
must be chosen so that we can carry out some or all of the integrations 
in Eq.(\ref{def-of-F-var}). We introduce variational parameters 
in the trial actions that are determined  by requiring that they minimize 
$\beta F_{\rm var}$. We have used two different choices for 
$S_0^{(\mu)}$. 

\subsection{ First variational calculation}

The first choice decouples all the phases in both arrays 
\begin{equation}
\label{first-choice-for-s0}
     S_0^{(\mu)}[\phi_\mu] = -\gamma_\mu\sum_{\vec r} \cos(\phi_\mu(\vec r)),
\end{equation}
with the two $\gamma_\mu$'s  the variational parameters
\cite{kleinert-book-2}. 
The advantage of this form for the trial action is that all the  integrals
can be analytically  computed  or they can be expressed as simpler
one--dimensional integrals. For the classical 2-D XY model
it is known that this variational choice  grossly overestimate
the critical temperature 
\cite{kleinert-book-2}. This approximation, nonetheless, gives good 
qualitative  results for  the critical temperatures for both JJA
layers. Here we are mostly interested in general trends 
so, for simplicity,  we only study the case when both JJA layers have 
the same parameter values. After a lengthy but direct calculation, the 
critical temperature equation for  one of arrays is obtained from solving
the equation
\begin{equation}
\label{first-choice-tc}
    D_4\ x^4+D_3\ x^3+D_2\ x^2+D_1\ x+D_0 = 0,
\end{equation}
were we found 
\begin{equation}
\label{def-of-x-and-alpha-d0-d1}
      x = \beta_c E_J,\ \ \ \ \alpha = e^2/(2C_{\rm s}E_J),\ \ \ \ \ 
      D_0 = -1,\ \ \ \ \ D_1 = 2,
\end{equation}
\begin{equation}
\label{def-of-d2-d3-d4}
      D_2 = -(4/3)\ \alpha K_1,\ \ \ \ \ \
      D_3 = \alpha^2 K_1^2/9,\ \ \ \ \ D_4 = 4\alpha^2(K_2/2-K_3)/45.
\end{equation}
$K_1, K_2$, and $K_3$ are complicated functions of the 
ratios $(C_{\rm int}/C_{\rm s})$ and $(C/C_{\rm s})$. They can
be computed in general in terms of summations over Fourier modes.  
Here we are only interested in a general trend of $T_c$ 
as a function of $C_{\rm {int}}$, and we further simplify the problem by 
considering the self-capacitive limit, i.e. $C_{\rm m}=0$. 
In this case, the  $K$ functions can be fully computed giving
\begin{equation}
\label{K1-for-Cm=0}
        K_1 = \left(\frac{1+C_{\rm int}/C_{\rm s}}{1+2C_{\rm int}
                C_{\rm s}}\right),\ \ \ \
        K_2 = 28K_1^2,\ \ \ \ and\ \ \ \
        K_3 = 8K_1^2.
\end{equation}
We can now interpret the effect of the inter-plane capacitance as a
rescaling  of the single layer quantum parameter $\alpha$, namely
\begin{equation}
\label{renorm-of-alpha}
         \alpha_{\rm eff} = \left(\frac{1+C_{\rm int}/C_{\rm s}}
                               {1+2C_{\rm int}/C_{\rm s}}\right)\alpha.
\end{equation}
The first conclusion we draw from this result is that, in the
semiclassical approximation, the inter-plane capacitance makes the  
system {\it less quantum mechanical}. The critical temperature
increases as  $(C_{\rm int}/C_{\rm s})$ increases up to an asymptotic plateau. 
To further check this result we have also performed quantum Monte Carlo 
calculations that confirmed our analytic results.

Note that here we only presented the case where both arrays are
equal and the mutual capacitance is zero. We have done so due to the 
simplicity of the analytic results. It is not much harder, within this 
approximation, to numerically calculate $T_c$ for the more general
cases. The general result leads to the same qualitative conclusion; The 
increase in the inter-plane capacitance raises the critical temperature 
for both arrays.

From Eq.(\ref{renorm-of-alpha}) we can find $T_c$ up to second order in 
the effective quantum parameter $\alpha_{\rm eff}$ giving
\begin{equation}
\label{tc-up-to-second-order}
      \left(\frac{k_BT_c}{E_J}\right) = \left(\frac{k_BT_c^{(0)}}{E_J}\right)
      - \left(\frac{2}{3}\right) \alpha_{\rm eff} + \left(\frac{189}{180}
        \right)\alpha_{\rm eff}^2 + O(\alpha_{\rm eff}^3),
\end{equation}
where the variational result for the classical 2-D XY model is
$(k_BT_c^{(0)}/E_J)=2$. It is evident that this approximation 
overestimated the critical temperature. Surprisingly, it gives a
very good estimate of the first two correction values for the one 
array problem \cite{jacobs}.

\subsection{Second variational calculation}

In the previous subsection we  used
Eq.(\ref{first-choice-for-s0}) to perform the MFT
 variational calculation. From the form of  
Eq.(\ref{def-of-F-var}) it is clear that it is not 
necessary to use a trial action that decouples all the phases in both 
arrays. We now want to use a better trial action that 
 gives better  2-D XY model classical results. This has the advantage
that the classical limit for each array is by construction  
exact. The disadvantage of this choice is that we need to evaluate a
nonlocal average over the 2-D XY model classical Hamiltonian,
for which  we have to approximately evaluate an infinite lattice sum.
Again, we restrict the calculation to the $C_{\rm m}=0$ limit, although 
we could do the numerical calculation for the full model.

The starting point of this scheme is to use the trial action
\begin{equation}
\label{second-choice-for-s0}
      S_0^{(\mu)}[\phi_\mu] = -\beta_\mu \sum_{<\vec r_1,\vec r_2>}
                \cos\left(\phi_\mu(\vec r_1)-\phi_\mu(\vec r_2)\right),
\end{equation}
with $\beta_1$, and $\beta_2$  the variational parameters needed to 
 calculate the variational free energy in Eq. 
(\ref{def-of-F-var}) from Eq.(\ref{eff-action-two-layers}).
We need to evaluate the following averages
\begin{eqnarray}
\label{average-for-cos}
   & & \left<F_\mu\right>_{XY} =\  E_J^{(\mu)} \sum_{<\vec r_1,\vec r_2>}
          \left(1-\Big<\cos\left(\phi_\mu(\vec r_1)-\phi_\mu(\vec r_2)\right)
                        \Big>_{XY}\right),\\
\label{average-for-cos-cos}
  & &\left<{\bf\tilde C}_{\mu,\nu}\ \frac{\partial^2 F_\nu}
           {\partial\phi_\nu^2}\ {\bf\tilde C}_{\nu,\mu}\ \frac{\partial^2
            F_\mu}{\partial\phi_\mu^2}\right>_{XY} \sim
            \Big< \cos\Big(\phi_\mu(\vec r_1)-\phi_\mu(\vec r_2)\Big)
                   \cos\Big(\phi_\nu(\vec r_3)-\phi_\nu(\vec r_4)\Big)
            \Big>_{XY}.\nonumber\\
\end{eqnarray}
The average in Eq.(\ref{average-for-cos}) is simple, since 
 we only need to perform the average when $\vec r_1$ and 
$\vec r_2$ are nearest neighbors. Here we are interested
in periodic and symmetric arrays.  It is then convenient to
define and evaluate the following short range correlation function
\begin{equation}
\label{def-of-g0}
        g_0(\beta) = \Big<\cos(\phi(\vec r)-\phi(\vec r+\vec u)\Big>_{XY},
\end{equation}
with $\beta$ is the inverse temperature of the classical 2-D XY model.
This function can not be calculated exactly, but it can be evaluated
using a Monte Carlo calculation or by matching a low to a high-temperature
expansion for the classical 2-D XY model.

The average in Eq.(\ref{average-for-cos-cos}) is more complicated.
First, if $\mu\ne\nu$, due to the intra-plane independence, the average
can be reduced to finding two values of $g_0(\beta)$. If $\mu=\nu$, 
the problem is intractable for general $\vec r_1$ and $\vec r_3$. This is
exactly the problem we would encounter if we want to make a calculation for a
general full capacitance matrix. On the other hand if we choose
 $C_{\rm m}=0$,  the problem is  simplified. In this case, only 
$g_0(\beta)$ and the following functions 
\begin{eqnarray}
\label{def-of-g1}
       g_1(\beta) &=& \Big<\Big[\cos\left(\phi(\vec r)-\phi(\vec r+\hat x)
                           \right)\Big]^2\Big>_{XY},\\
\label{def-of-g2}
       g_2(\beta) &=& \Big<\cos\left(\phi(\vec r)-\phi(\vec r+\hat x)\right)
                           \cos\left(\phi(\vec r)-\phi(\vec r+\hat y)\right)
                      \Big>_{XY},\\
\label{def-of-g3}
       g_3(\beta) &=& 2\left\{\frac{2(dg_1/d\beta)+(dg_2/d\beta)}{dg_0/d\beta}
                       \right\},
\end{eqnarray}
need to be known.
These function can be calculated using a Monte Carlo calculation for the
classical 2-D XY model. Using all these functions, we can evaluate the
variational free energy $F_{\rm var}(\beta_1,\beta_2)$. The parameters
$\beta_1$ and $\beta_2$ are found by imposing the condition that they 
minimize the free energy. This condition reads
\begin{equation}
\label{eq-for-1}
       \frac{\partial F_{\rm var}}{\partial\beta_1} =0,\,\,\,\,
       \frac{\partial F_{\rm var}}{\partial\beta_2} = 0.
\end{equation}
The explicit equation for $\beta_1$ obtained from Eq.(\ref{eq-for-1}) is
\begin{eqnarray}
\label{explicit-from-eq-for-1}
      \beta_1 = \!\!\!\!\!
                & &(\beta E_J^{(1)})\Bigg\{1-\frac{(\beta E_J^{(1)})}{3}
                   \alpha_{\rm eff}^{(1)}+\frac{(\beta E_J^{(1)})^2}{18}
                   {\alpha_{\rm eff}^{(1)}}^2\Bigg\} 
                +  \frac{4(\beta E_J^{(1)})^4}{45}\Bigg\{
                   {\alpha_{\rm eff}^{(1)}}^2 g_3(\beta_1) +\nonumber\\
               &+& 16\left(\frac{E_J^{(2)}}{E_J^{(1)}}\right)^2
                   \alpha_{\rm eff}^{(1)} \alpha_{\rm eff}^{(2)}
                   \left(\frac{C_{\rm int}/C_{\rm s}^{(1)}}
                              {1+C_{\rm int}/C_{\rm s}^{(1)}}\right)
                   \left(\frac{C_{\rm int}/C_{\rm s}^{(2)}}
                              {1+C_{\rm int}/C_{\rm s}^{(2)}}\right)
                   g_0(\beta_2)\Bigg\}.
\end{eqnarray}
A similar equation can be written down for $\beta_2$.
In writing this equation we have  used the following 
definition for the effective quantum parameter for the  arrays
\begin{equation}
\label{def-of-alpha-mu-eff}
      \alpha_{\rm eff}^{(1)} = \frac{e^2}{2C_{\rm s}^{(1)}E_J^{(1)}}
      \left(\frac{1+C_{\rm int}/C_{\rm s}^{(2)}}{1+C_{\rm int}/C_{\rm s}^{(1)}
            +C_{\rm int}/C_{\rm s}^{(2)}}\right).
\end{equation}
 We are therefore left with the following set of self-consistent
equations for the two  variational parameters
\begin{equation}
\label{eq-for-beta1}
                      \beta_1 = G_1(\beta,\beta_1,\beta_2),\,\,\,\,
                      \beta_2 = G_2(\beta,\beta_1,\beta_2).
\end{equation}
The functions $G_1$ and $G_2$ can be identified from Eq. 
(\ref{explicit-from-eq-for-1}).
We set $\beta_\mu=\beta_c^{XY}$ as the condition to find the critical 
temperature of the array $\mu$, i.e. we identify the value
of  $\beta_\mu$ with an effective inverse temperature
for the array $\mu$. To find the
critical temperature for array 1 we solve the following set of equations
\begin{equation}
\label{eq-for-tc1}
                  \beta_c^{XY} = G_1(\beta_c^{(1)},\beta_c^{XY},\beta_2),
\,\,\,\,
                  \beta_2\   = G_2(\beta_c^{(1)},\beta_c^{XY},\beta_2).
\end{equation}
These equations are not difficult to solve. First, given a value of
 $\beta$ we first solve for $\beta_2$ using Eq.(\ref{eq-for-tc1}).
Note that this equation can be written in the following form
\begin{equation}
\label{eq-to-get-beta2}
          \beta_2 = A + B\ g_3(\beta_2).
\end{equation}
We can show from Eq.(\ref{explicit-from-eq-for-1})  that the
quantities $A$ and $B$ in this equation are not functions of
$\beta_2$. Moreover, solving this 
equation using a fixed point search method gives 
$\beta_2(\beta_c^{(1)})$, that can then be introduced into Eq. 
(\ref{eq-for-tc1}). We are now left with a one variable equation.

The case when both arrays have the same parameters is again easier
to solve, since the transcendental Eqs. (\ref{eq-for-tc1})  
reduce to just one polynomial equation
\begin{equation}
\label{polinomial-eq-for-tc}
     R_4 x^4 + R_3 x^3 + R_2 x^2 + R_1 x + R_0 = 0,
\end{equation}
where $\beta_c^{(1)}=\beta_c^{(2)}$ and
\begin{equation}
\label{def-of-r0-r1-r2-r3}
   R_0 = -\beta_c^{XY},\ \ \ \ \ R_1 = 1,\ \ \ \ R_2 = -\frac{2}{3}
   \alpha_{\rm eff},\ \ \ \ R_3 = \frac{1}{18}\alpha_{\rm eff}^2,
\end{equation}
\begin{equation}
\label{def-of-r4}
   R_4 = \frac{4}{45}\alpha_{\rm eff}^2\left[3.6+9.6\left(
         \frac{C_{\rm int}/C_{\rm s}}{1+C_{\rm int}/C_{\rm s}}\right)^2\right].
\end{equation}
We have used Eq.(\ref{renorm-of-alpha}) to define $\alpha_{\rm eff}$
and found the value of $g_3(\beta_c^{XY})$ using Monte Carlo calculations 
to be
\begin{equation}
\label{def-beta-xy}
       \beta_c^{XY} \approx 1.1186,\,\,\,\,
        g_3(\beta_c^{XY}) \approx 3.6\ .
\end{equation}
The result of this calculation gives the same result as in  
Eq.(\ref{tc-up-to-second-order}) for the first order correction
in $\alpha$, but now the $\alpha=0$ limit gives 
the correct classical value for $T_c/k_BE_J=1/\beta_c^{XY}$.
Here the general result is qualitatively the same; an
increase in the interaction capacitance $C_{\rm int}$ results
in a decrease of the effective quantum parameter.

%
%
\section{Duality in two capacitively coupled JJA}
\label{sec:quantum-limit}
In the previous section we studied the specific changes in
the individual critical temperature of two JJA, when both were
in the semiclassical parameter regime. In this section we
analytically consider the interesting case where one array is in
the semiclassical regime and the other in the quantum one. In the
semiclassical array,  vortices are localized while the Cooper pair 
charge fluctuations are mobile in the superfluid phase. This happens for small
$\alpha$ values. In the quantum JJA regime, large $\alpha$,
 the vortices are mobile while
the charges are localized. It is not possible to have both vortices and
charges simultaneously localized or mobile since this is forbidden by the
Heisenberg uncertainty principle (which has been shown at work in recent
experiments  in a small Josephson array system \cite{heisenberg}).
The interaction between vortices and charges has a minimal coupling
form, with  constant strength and it is sharply localized, {\sl i.e.}
a vortex and a Cooper pair only interact if they are located at the
same point in the array. By considering two arrays with one vortex
dominated and the other charge dominated we can have them both
interacting via the coupling capacitance between the arrays.
A related analysis of the coupled array system was considered in Ref.
\cite{meeting} by us and an alternative and complementary 
analysis was also presented in Ref. \cite{blatter}.

In this section we will carry out the two array analysis extending techniques 
developed for the study of one array \cite{fazio-schon-91}.
Here we shall consider first the one-array component of the Hamiltonian 
given in  Eq.(\ref{hamiltonian-with-c-tilde}) and later its Villain 
approximation \cite{jkkn}. We briefly mention  the one JJA 
calculational approach, since the extension to the two-array case follows 
from this analysis. This is true only because the arrays are 
electrostatically coupled.

The Hamiltonian for one array reads
\begin{equation}
\label{ham-for-one-array}
      {\hat H} = \frac{q^2}{2} \sum_{\vec r_1,\vec r_2} \hat n(\vec r_1)
                {\bf C}^{-1}(\vec r_1,\vec r_2) \hat n(\vec r_2)
                +E_J\sum_{<\vec r_1,\vec r_2>} \Big[1-\cos\left(
                 \hat\phi(\vec r_1)-\hat\phi(\vec r_2)\right)\Big],
\end{equation}
where ${\bf C}(\vec r_1,\vec r_2)$ is the one-array capacitance.
The partition function for this Hamiltonian can be written in the 
path integral form
\begin{eqnarray}
\label{part-func-for-one-array}
 Z = \prod_{\vec r} \prod_{\tau} \sum_{\{n(\tau,\vec r)\}} \int_{0}^{2\pi}
     & &
     \frac{d\phi(\tau,\vec r)}{2\pi} \exp\Bigg\{-\int_{0}^{\beta\hbar}
      d\tau\Bigg[\frac{q^2}{2}\sum_{\vec r_1,\vec r_2}n(\tau,\vec r_1)
      {\bf C}^{-1}(\vec r_1,\vec r_2)n(\tau,\vec r_2)+\nonumber\\
   & & + i \sum_{\vec r} n(\tau,\vec r)\frac{d\phi}{d\tau}(\tau,\vec r)
      +E_J\!\!\!\sum_{<\vec r_1,\vec r_2>} \Big(1-\cos\phi_{\vec r_1,\vec r_2}
      (\tau)\Big)\Bigg]\Bigg\}.
\end{eqnarray}
Here we denote
$\phi_{\vec r_1,\vec r_2}(\tau) = \phi(\tau,\vec r_1)-\phi(\tau,\vec r_2)$,
 and write the imaginary time summation as an integral. To  integrate 
over the $\phi$'s,  we  need to introduce an additional set of variables. 
This is  done  by writing the Boltzmann factor as a Fourier series using
the Poisson summation representation \cite{jkkn}, 
\begin{eqnarray}
\label{fourier-series-2}
   & &\exp\left\{-\lambda\left(1-\cos\psi\right)\right\} = 
      \sum_{m=-\infty}^{\infty} f_m(\lambda)\ \exp\{i\psi m\},\\
\label{fourier-coefs}
   & &\!\!\!\!\!\! f_m(\lambda) = \int_{0}^{2\pi} \frac{d\psi}{2\pi} 
      \exp\{-\lambda(1-\cos\psi)\}\ \exp\{-i\psi m\} = 
      \exp\{-\lambda\}\ I_m(\lambda).
\end{eqnarray}
Here  $I_m(\lambda)$ is the modified Bessel function. The integral in
Eq.(\ref{fourier-coefs}) gives a convenient way to extract the
asymptotics for the small and large $\lambda$. For small $\lambda$, the
Taylor series expansion of $I_m(\lambda)$ around $\lambda=0$ 
gives us the leading expansion terms of
$f_m(\lambda)$. For large $\lambda$, a steepest descent calculation
yields the leading term. In these two asymptotic limits we have
\begin{equation}
\label{small-large-lambda}
     f_m(\lambda) \approx \left\{ \begin{array}{ll}
                                     (\lambda/2)^m/m!, 
                                         & \mbox{\ \ if $\lambda\ll 1$},\\
                                         & \\
                                     \exp\{-m^2/2\lambda\}/\sqrt{2\pi\lambda},
                                          & \mbox{\ \ if $\lambda\gg 1$}.
                                   \end{array}
                          \right.
\end{equation}
If we use the $\lambda\gg 1$ result  in Eq.(\ref{fourier-series-2}), 
we can write
\begin{equation}
\label{exp-cos-for-large-lambda}
          \exp\left\{-\lambda\left(1-\cos\psi\right)\right\} \approx
          \frac{1}{\sqrt{2\pi\lambda}} \sum_{m=-\infty}^{\infty}
          \exp\{-m^2/2\lambda+i\psi m\}.
\end{equation}
Using Eq.(\ref{fourier-series-2}) in Eq.(\ref{part-func-for-one-array}) 
we have new summation link variables $m_\nu(\tau,\vec r)$ 
between the nodes of the lattice. This is repeated for each of the 
imaginary time planes. The subindex $\nu$ is a vector that
denotes the orientation of the links so that we can write
\begin{eqnarray}
\label{epprox-for-exp-cos}
   \exp\Big\{-\epsilon E_J
   & &\Big[1-\cos\Big(\phi(\tau,\vec r+\hat\nu)-
   \phi(\tau, \vec r)\Big)\Big]\Big\} \approx
   \left(\frac{1}{2\pi\epsilon E_J}\right)^{L_\tau L_x L_y} \times\nonumber\\
   & &\times\sum_{m_\nu(\tau,\vec r)}\ \exp\Bigg\{
   -\frac{1}{2\epsilon E_J}m_\nu(\tau,\vec r)^2 +
        i m_\nu(\tau,\vec r)\Delta_\nu\phi(\tau,\vec r)\Bigg\}.
\end{eqnarray}
Here we have discretized the imaginary time interval with
$\epsilon=\beta\hbar/L_\tau$, and the lattice derivative is 
$\Delta_\nu f(\vec r) = f(\vec r+\hat\nu)-f(\vec r)$. After the 
 integrations over
the $\phi$'s the result can be written
 in terms of the integer $n$ and $m$ variables as
\begin{equation}
\label{z-after-int-phis}
    Z \approx \!\!\sum_{\{n(\tau,\vec r)\}} {\sum_{\{\vec m(\tau,\vec r)\}}}
    \!\!\!\exp\Bigg\{\!\!-\!\!\sum_{\tau}\Bigg[\frac{\epsilon q^2}{2}
    \sum_{<\vec r_1,\vec r_2>}\!\!n(\tau, \vec r_1){\bf C}^{-1}
    (\vec r_1,\vec r_2)n(\tau,\vec r_2)+
    \frac{1}{2\epsilon E_J}\sum_{\vec r} |\vec m(\tau,\vec r)|^2\Bigg]
     \Bigg\}.
\end{equation}
After the integration over the $\phi$'s we get a set of constraints 
over the $n$ and $m$ values. These constraints can be written as 
discrete continuity equations that are satisfied at each node of the array,
\begin{equation}
\label{cont-equation}
   \vec\Delta\cdot\vec m(\tau,\vec r) + \Delta_\tau n(\tau,\vec r) = 0.
\end{equation}
These constraint equations can be solved in several ways. For example, the 
pair $(n,\vec m)$ can be expressed in terms of a 
three-vector ${\cal K}=(n,\vec m)$ \cite{stern}, so that 
 Eq.(\ref{cont-equation}) becomes
\begin{equation}
\label{eq-in-three-vector-notation}
       \Delta_\nu {\cal K}^\nu = 0,
\end{equation}
with the discrete gradient, $\Delta_\nu=(\Delta_\tau,\vec{\Delta})$. 
From  Eq.(\ref{eq-in-three-vector-notation})  ${\cal K}$ can 
be expressed as the curl of a gauge field, i.e.
${\cal K}^\mu = \varepsilon^{\mu\nu\rho}\Delta_\nu {\cal A}_\rho$,
with  $\varepsilon^{\mu\nu\rho}$ the usual fully antisymmetric tensor.
Substituting this result into Eq.(\ref{z-after-int-phis}) we get
an effective action over the gauge field ${\cal A}$, which
resolves the constraints over the summations. 

Our solution to the constraint in Eq.(\ref{cont-equation})
is different from the one used in  Ref. \cite{fazio-schon-91} where 
they wanted to preserve the  $n$ variables. The solution to our constraint 
equation  will have  a particular solution plus an homogeneous solution. Note that
Eq.(\ref{cont-equation}), written in this form,  resembles  
one of Maxwell's equation that connect the divergence of the electric field 
to the charge density. The particular solution to this equation 
contains the  gradient of a line integral, that can then be solved 
using a discrete line integral operator. The solution can be formally 
written as
\begin{equation}
\label{solution-to-the-constraint}
       m^{\mu}(\tau,\vec r) = -e^\mu(\hat e\cdot\vec\Delta)^{-1}
       \Delta_\tau n(\tau,\vec r) + \varepsilon^{\mu\nu} \Delta_{\nu} 
       A(\tau,\vec r),
\end{equation}
with $A(\tau,\vec r)$ another integer gauge field. To obtain the 
partition function we need to perform summations over this field. The 
first term in the last equation represents a discrete line integral, 
which is better calculated in a Fourier representation. 
After substituting Eq.(\ref{solution-to-the-constraint}) into 
Eq.(\ref{z-after-int-phis}) we obtain an expression for an effective 
action as functions of the $n$'s and the $A$'s. The partition function 
is now obtained by summing over these variables, but now 
without any constraint. 
We perform the summation over the $A$'s again with the help of the 
Poisson summation formula. After  introducing
a new set of integer $v$ variables, the summation over the $A$'s can now
be done since the integrals left to calculate are
Gaussian and the $A$'s are unconstrained fields. The final result is
\begin{eqnarray}
\label{Z-with-ns-and-vs}
    Z &=& \sum_{\{n\}}\sum_{\{v\}} \exp\Bigg[-S_{\rm eff}(n,v)\Bigg],\\
\label{def-of-seff-with-n-and-v}
    S_{\rm eff}(n,v) &=& \sum_{\vec r_1,\vec r_2,\tau}\Bigg[
    \frac{q^2\epsilon}{2\pi} n(\tau,\vec r_1){\bf C}(\vec r_1,\vec r_2)
    n(\tau,\vec r_2)+\pi\epsilon E_J v(\tau,\vec r_1){\bf G}(\vec r_1,
    \vec r_2)v(\tau,\vec r_2)+\nonumber\\
    & &+i\ n(\tau,\vec r_1){\bf\Theta}(\vec r_1,\vec r_2)
    \Delta_\tau v(\tau,\vec r_2)+\frac{1}{4\pi\epsilon E_J} \Delta_\tau
    n(\tau,\vec r_1){\bf G}(\vec r_1,\vec r_2)\Delta_\tau n(\tau,\vec r_2)
    \Bigg],\nonumber\\
\end{eqnarray}
where we defined 
\begin{equation}
\label{def-of-G}
     {\bf G}(\vec r_1,\vec r_2) \approx \ln|\vec r_1-\vec r_2|,
     \,\,\,\,\, and \,\,\,
{\bf\Theta}(\vec r_1,\vec r_2) \approx \arctan\left(\frac{y_1-y_2}
     {x_1-x_2}\right).
\end{equation}
Eq.(\ref{def-of-seff-with-n-and-v}) is  an effective action for two  
coupled imaginary time Coulomb gases. This equation is valid for 
all parameter ranges. When
$E_J$ is large, the last term in this equation is small and 
the time derivatives of the charges are soft with the  $n$'s having strong
fluctuations.  In this limit the $v$'s dominate and the $n$'s are not well 
defined. We will call this a vortex-dominated regime. When $E_J$ is small, 
the last term in the effective action is large and it  makes the 
time derivatives  of the charges  well defined. In
this limit the $v$'s are not well defined  and the state is 
charge-dominated,
with the charges  
described by an effective continuous Gaussian model. After integrating
the continuous variables 
we obtain an effective action for the vortex integer conjugate variables.

One important aspect of the one-array derivation of 
Eq.(\ref{def-of-seff-with-n-and-v}) 
is that it did not involve  
the charging energy part. This means that when we carry out the 
two-array calculations   we only need to see that
 its  effective action can be written down immediately
from just repeating the one array calculation; we only need to
add the essential extra charging energy term that couples the two arrays. 
The two-array equivalent equation to Eq.(\ref{def-of-seff-with-n-and-v}) 
is then
\begin{eqnarray}
\label{eff-action-for-two-arrays}
     S_{\rm eff}(n^{(1)},v^{(1)};n^{(2)},v^{(2)}) =& &
     \sum_{\vec r_1,\vec r_2,\tau} \Bigg[\frac{q^2\epsilon}{2\pi}
     n^{(1)}(\tau,\vec r_1){\bf\tilde C}_{1,1}(\vec r_1,\vec r_2) n^{(1)}
     (\tau,\vec r_2)+\nonumber\\
   & &\ \ \ \ \ \ \ \ \ \ +\pi\epsilon E_J^{(1)} v^{(1)}(\tau,\vec r_1){\bf G}
     (\vec r_1,\vec r_2) v^{(1)}(\tau,\vec r_2)+\nonumber\\
   & &\ \ \ \ \ \ \ \ \ \ +i\ n^{(1)}(\tau,\vec r_1){\bf\Theta}
     (\vec r_1,\vec r_2)\Delta_\tau v^{(1)}(\tau,\vec r_2)+\nonumber\\
   & &\ \ \ \ \ \ \ \ \ \ +\frac{1}{4\pi\epsilon E_J^{(1)}}\Delta_\tau 
      n^{(1)}(\tau,\vec r_1){\bf G}(\vec r_1,\vec r_2)
      \Delta_\tau n^{(1)}(\tau,\vec r_2)\Bigg]+\nonumber\\
   & &+ \sum_{\vec r_1,\vec r_2,\tau} \Bigg[\frac{q^2\epsilon}{2\pi}
     n^{(2)}(\tau,\vec r_1){\bf\tilde C}_{1,1}(\vec r_1,\vec r_2) n^{(2)}
     (\tau,\vec r_2)+\nonumber\\
   & &\ \ \ \ \ \ \ \ \ \ \ +\pi\epsilon E_J^{(2)} v^{(2)}(\tau,\vec r_1)
      {\bf G}
     (\vec r_1,\vec r_2) v^{(2)}(\tau,\vec r_2)+\nonumber\\
   & &\ \ \ \ \ \ \ \ \ \ \ +i\ n^{(2)}(\tau,\vec r_1){\bf\Theta}
      (\vec r_1,\vec r_2)\Delta_\tau v^{(2)}(\tau,\vec r_2)+\nonumber\\
   & &\ \ \ \ \ \ \ \ \ \ \ +\frac{1}{4\pi\epsilon E_J^{(2)}}
      \Delta_\tau n^{(2)}(\tau,\vec r_1){\bf G}(\vec r_1,\vec r_2)
      \Delta_\tau n^{(2)}(\tau,\vec r_2)\Bigg]+\nonumber\\
   & & +\sum_{\vec r_1,\vec r_2,\tau} 
       \Bigg[\frac{q^2\epsilon}{\pi}n^{(1)}(\tau,\vec r_1){\bf\tilde C}
       _{1,2}(\vec r_1,\vec r_2) n^{(2)}(\tau,\vec r_2)\Bigg],
\end{eqnarray}
where we used the definition of ${\bf\tilde C}$ given in Eq. 
(\ref{def-of-c-tilde}). 

In this section we shall consider  the interesting case where one of the arrays
is in the semiclassical regime (the vortex-dominated state) and the 
other is in the full quantum regime (i.e. the charge-dominated state). We want
in particular to study the interaction between vortices in 
one array and charges in the other. We assume, as in experiment, that the 
arrays are dominated by the mutual capacitance between nearest neighbors.
We take array $1$  vortex-dominated and array $2$  
charge-dominated, i.e.
\begin{equation}
\label{inequality-for-array1}
      E_{C_{\rm m}}^{(1)} \ll E_J^{(1)},\,\,\,\,\,
      E_{C_{\rm m}}^{(2)} \gg E_J^{(2)}.     
\end{equation}
We start by performing a vortex integration in array $2$.
This can be done  using the Poisson summation formula to write
\begin{eqnarray}
\label{poisson-for-array2}
     \sum_{\{v^{(2)}\}}\exp\Bigg[-S^{(2)}(v^{(1)})\Bigg] =
     \sum_{\{P\}}\int& &\prod_{\tau,\vec r} d\Phi(\tau,\vec r)
     \exp\Big[-S^{(2)}(\Phi)\Big]\times\nonumber\\
 & & \times\exp\Bigg[2\pi i\sum_{\tau,\vec r}\Phi(\tau,\vec r) 
     P(\tau,\vec r)\Bigg].
\end{eqnarray}
We can neglect the $P\neq 0$ terms when $E_{C_{\rm m}}^{(2)} \gg E_J^{(2)}$, 
since they are exponentially small. Note that this integration only 
involves vortices  in array $2$, and it can 
be done without affecting the variables in array $1$. 
When we only consider the $P=0$ terms, there is a change in the
Josephson coupling constant given by $E_J^{(2)}\rightarrow E_J^{(2)}/2$
 \cite{fazio-schon-91}. In this parameter limit we see 
that the only modification is in the new coupling constant 
in Eq.(\ref{eff-action-for-two-arrays}).

We can also use the Poisson summation formula for the integration of the 
charges in array $1$. After the integration, the action for 
the charges can be written as
\begin{equation}
\label{action-for-charges-in-array1}
    S^{(1)}[n^{(1)}] = \frac{1}{2} \sum_{\tau,\tau',\vec r_1,\vec r_2}
    n^{(1)}(\tau,\vec r_1){\bf M}(\tau,\tau';\vec r_1,\vec r_2) n^{(1)}
    (\tau',\vec r_2)+\sum_{\tau,\vec r} J(\tau,\vec r) n^{(1)}(\tau,\vec r),
\end{equation}
where  the ${\bf M}$ operator and effective
current $J$ are 
\begin{eqnarray}
\label{def-of-M}
  & &  {\bf M}(\tau,\tau';\vec r_1,\vec r_2) = \frac{q^2\epsilon}{\pi}
       {\bf\tilde C}_{1,1}(\vec r_1,\vec r_2) \delta_{\tau,\tau'} -
       \frac{1}{2\pi\epsilon E_J^{(1)}}{\bf G}(\vec r_1,\vec r_2) 
       \Delta_\tau^2,\\
\label{def-of-j}
 & &   J(\tau,\vec r_1) = \sum_{\vec r_2}\Bigg[i{\bf\Theta}(\vec r_1,\vec r_2)
       \Delta_\tau v^{(1)}(\tau,\vec r_2)+\frac{q^2\epsilon}{\pi}
       {\bf\tilde C}_{1,2}(\vec r_1,\vec r_2) n^{(2)}(\tau,\vec r_2)\Bigg].
\end{eqnarray}
After integrating the charges in array 1 and the vortices in array 2 we 
are left with the following expression for the effective partition function
\begin{equation}
\label{z-with-n1-v2}
     Z = \sum_{\{v^{(1)}\}}\sum_{\{n^{(2)}\}} \exp\Big[-S_{\rm eff}(v^{(1)},
     n^{(2)})\Big],
\end{equation}
where the effective action for vortices in array 1 and charges in array 2 is
given by
\begin{eqnarray}
\label{seff-for-v1-n2}
 S_{\rm eff}[v^{(1)},n^{(2)}] & & =\sum_{\tau,\vec r_1,\vec r_2}
      \Bigg[\pi\epsilon E_J^{(1)}\ v^{(1)}(\tau,\vec r_1){\bf G}(\vec r_1,
      \vec r_2)\ v^{(2)}(\tau,\vec r_2)+\nonumber\\
  & & \ \ \ \ \ \ \ \ \ +\frac{1}{2\pi\epsilon E_J^{(2)}}\Delta_\tau n^{(2)}
       (\tau,\vec r_1){\bf G}(\vec r_1,\vec r_2)\Delta_\tau n^{(2)}
       (\tau,\vec r_2)\Bigg]+\nonumber\\
  & & +\sum_{\tau,\tau',\vec r_1,\vec r_2}\Bigg[\frac{\epsilon}{2}
       n^{(2)}(\tau,\vec r_1){\bf G_n}(\tau,\tau';\vec r_1,\vec r_2)
       n^{(2)}(\tau',\vec r_2)+\nonumber\\
  & & \ \ \ \ \ \ \ \ \ \ \ \ \ +i\ n^{(2)}(\tau,\vec r_1){\bf\tilde\Theta}
       (\tau,\tau';\vec r_1,\vec r_2)\Delta_\tau v^{(1)}
       (\tau',\vec r_2)+\nonumber\\
  & & \ \ \ \ \ \ \ \ \ \ \ \ \ +\frac{\pi}{2q^2\epsilon}\Delta_\tau v^{(1)}
       (\tau,\vec r_1){\bf G_v}(\tau,\tau';\vec r_1,\vec r_2)
       \Delta_\tau v^{(1)}(\tau',\vec r_2)\Bigg].
\end{eqnarray}
The effective interaction potentials are defined by
\begin{eqnarray}
\label{def-of-Gn}
      {\bf G_n}(\tau,\tau';\vec r_1,\vec r_2) &=& \frac{q^2}{\pi}
      \Bigg[{\bf\tilde C}_{2,2}(\vec r_1,\vec r_2)\delta_{\tau,\tau'}-
      \nonumber\\ & & \  -
      \frac{q^2\epsilon}{\pi}\sum_{\vec r_3,\vec r_4} {\bf\tilde C}_{1,2}
      (\vec r_1,\vec r_3)
      {\bf M}^{-1}(\tau,\tau';\vec r_3,\vec r_4){\bf\tilde C}_{1,2}
      (\vec r_4,\vec r_2)\Bigg],\\
\label{def-of-theta-tilde}
      {\bf\tilde\Theta}(\tau,\tau';\vec r_1,\vec r_2) &=& -
      \frac{q^2\epsilon}{\pi}\sum_{\vec r_3,\vec r_4} 
      {\bf\Theta}(\vec r_2,\vec r_3) {\bf M}^{-1}(\tau,\tau';\vec r_3,\vec r_4)
      {\bf\tilde C}(\vec r_4,\vec r_2),\\
\label{def-of-Gv}
      {\bf G_v}(\tau,\tau';\vec r_1,\vec r_2) &=& \sum_{\vec r_3,\vec r_4}
      {\bf\Theta}(\vec r_1,\vec r_3){\bf M}^{-1}(\tau,\tau';\vec r_3,\vec r_4)
      {\bf\Theta}(\vec r_4,\vec r_2).
\end{eqnarray}
Notice that the time nonlocality of these kernels comes from the second 
term in Eq.(\ref{def-of-M}). To gain some physical understanding of these 
complicated equations we  will next discuss a simplification 
when the nonlocal term in Eq.(\ref{def-of-M}) is small.

\subsection{\bf Vortex-charge capacitive gauge-like coupling}

In Eq.(\ref{seff-for-v1-n2}) we have an effective interaction between
vortices in array $1$ and charges in array $2$. We consider the dynamics 
of just one charge and one vortex in each array, in a similar way as 
was done for one array in Ref. \cite{fazio-schon-91}. Lets assume 
that the vortex and  the charge move along the imaginary time-dependent 
trajectories $\vec R(\tau)$, and $\vec X(\tau)$ respectively. In this case the 
vortex and charge space-time distributions can be described by
\begin{equation}
\label{one-vortex}
       v^{(1)}(\tau,\vec r) = \delta[\vec r-\vec R(\tau)];\, \, \, \, \, \, 
\, \, \, \, \, \, n^{(1)}(\tau,\vec r) = \delta[\vec r-\vec X(\tau)].
\end{equation}
After taking the time derivative of $v^{(1)}$ we find
\begin{equation}
\label{time-derivative-for-one-vortex}
      \Delta_\tau v^{(1)}(\tau,\vec r) = \Delta_\tau
      \delta[\vec r_1-\vec R(\tau)] = -\sum_{\mu} \Delta_\mu\delta[\vec r_1
      -\vec R(\tau)]\ \Delta_\tau\vec R(\tau).
\end{equation}
The right hand side in this equation relates the time derivative 
to a summation over space derivatives. We can next rewrite the interaction 
term in the  effective action in the following way
\begin{equation}
\label{int-action-xr}
    S_{\rm int} = -i\int_0^{\beta\hbar} d\tau \int_0^{\beta\hbar}
    d\tau' \sum_{\mu} \Delta_\mu {\bf\tilde\Theta}\left(\tau,\tau';
    \vec X(\tau),\vec R(\tau')\right)\ \frac{dR_\mu(\tau')}{d\tau}.
\end{equation}
This expression has a similar form to the typical minimal gauge coupling
in electrodynamics. Following this analogy we can define the
corresponding  vector potential
\begin{equation}
\label{def-of-a}
   \vec A(\tau,\vec r) = \int_0^{\beta\hbar} d\tau \vec\Delta
        {\bf\tilde\Theta}\left(\tau,\tau';\vec r,\vec X(\tau)\right).
\end{equation}
Using this definition Eq.(\ref{int-action-xr}) can be rewritten as
\begin{equation}
\label{new-xr-int-action}
    S_{\rm int} = -i\int_0^{\beta\hbar} d\tau'\ \vec A\left(\tau',
        \vec R(\tau')\right)\cdot\frac{d\vec R(\tau')}{d\tau}.
\end{equation}
Here we have chosen to view the vortex-charge interaction in the 
representation where the vortex moves under the influence of 
the charge gauge-like field $\vec A$. This view is equivalent to 
the representation where the charge  moves in the 
gauge-like field produced by the vortex. We have a vector field, and
we can find its corresponding effective ``magnetic field'' 
\begin{eqnarray}
\label{effective-magnetic-field}
    \vec B(\tau,\vec r) &=& \vec\Delta\times\vec A(\tau,\vec r),\nonumber\\
    &=& -\frac{q^2}{\pi} \int_0^{\beta\hbar} 
     d\tau' \sum_{\vec r_3,\vec r_4}\left(
    \vec\Delta\times\vec\Delta{\bf\Theta}(\vec X(\tau'),\vec r_3)\right)
    {\bf M}^{-1}(\tau,\tau';\vec r_3,\vec r_4) {\bf\tilde C}_{1,2}
    (\vec r_4,\vec r).\nonumber\\
\end{eqnarray}
The solution kernel  ${\bf\Theta}$ for a point vortex at the origin
satisfies the equation
\begin{equation}
\label{eq-for-theta}
       \vec\Delta\times\vec\Delta{\bf\Theta}(\vec r_1,\vec r_2) =
       2\pi\ \delta_{\vec r_1,\vec r_2}\ \hat{k},
\end{equation}
from which we can get our final expression for the effective magnetic field
\begin{equation}
\label{final-mag-field}
     \vec b(\tau,\vec r) = -\frac{q^2\epsilon}{\pi}\int_0^{\beta\hbar}
     d\tau' \sum_{\vec r_1} {\bf M}^{-1}(\tau,\tau';\vec X(\tau'),\vec r_1)
     {\bf\tilde C}_{1,2}(\vec r_1,\vec r)\ \hat k.
\end{equation}
Up to now we have that the effective action in Eq.(\ref{seff-for-v1-n2}), the
effective gauge vector potential, and its magnetic field interaction
are nonlocal in time, due to  
the nonlocality in time of the second term in Eq. 
(\ref{def-of-M}). In the limit $(\beta E_J^{(1)})(\beta E_{C_{\rm
m}}^{(1)})\gg 1$, the second term in Eq.(\ref{def-of-M}) is negligible,
and therefore we can write 
\begin{equation}
\label{limit-for-inv-m}
     {\bf M}^{-1}(\tau,\tau';\vec r_1,\vec r_2) \approx 
     \frac{\pi}{q^2\epsilon} {\bf\tilde C}_{1,1}^{-1}(\vec r_1,\vec r_2)\ 
     \delta_{\tau,\tau'}.
\end{equation}
Using Eq.(\ref{c-tilde-11}) we can write this equation in terms of 
the intra-array capacitance matrix
\begin{equation}
\label{C11-with-C1-C2}
       {\bf\tilde C}_{1,1}^{-1} = [{\bf C}_1-C_{\rm int}^2{\bf C}_2^{-1}].
\end{equation}
We note  that the array is periodic and symmetric, so that all
the commuting matrix operators can be diagonalized using plane waves. 
These facts produce important simplifications in the rest of the interaction
kernels, giving the results
\begin{eqnarray}
\label{limit-for-theta}
  & & {\bf\tilde\Theta} \approx -C_{\rm int}\ {\bf\Theta} {\bf C}_2^{-1}\ 
                          \delta_{\tau,\tau'},\\
\label{limit-for-Gn}
  & & {\bf G_n} \approx \frac{q^2}{\pi}\ {\bf C}_2^{-1}\ \delta_{\tau,\tau'}
                        ,\\
\label{limit-for-Gv}
  & & {\bf G_v} \approx {\bf\Theta} [{\bf C}_1-C_{\rm int}{\bf C}_2^{-1}]
                        {\bf\Theta}\ \delta_{\tau,\tau'}.
\end{eqnarray}
Eq.(\ref{limit-for-Gn}) is particularly significant, since it implies
that, within this approximation,  the interactions among  charges
in array $2$ do not depend on the presence of array $1$. This is a
counter-intuitive result, because we would expect that a virtual
photon excited from an island in array 2 and absorbed in another 
island in the same array would have contributions from bounce interactions with
array $1$.
What happens is that after adding all the contributions from these bounces, 
the net result (within the approximation leading to Eq.(\ref{limit-for-inv-m}))
is a cancellation of the contributions arising from array 1.
Finally, we can write the effective magnetic field as
\begin{equation}
\label{limit-for-beff}
       \vec b(\vec r) \approx -C_{\rm int}\ {\bf C}_2^{-1}
                              (\vec r,\vec X(\tau))\ \hat k.
\end{equation}
This result implies that, if we have a charge at $\vec X(\tau)$ and a vortex at
$\vec R(\tau)$, the vortex will feel an effective magnetic field
produced by the charge of magnitude 
$-C_{\rm int} {\bf C}_2^{-1}(\vec X(\tau),\vec R(\tau))$. This
situation is reminiscent of the vortex-charge bound states
extensively studied in the fractional quantum Hall effect problem,
and it may very well be that this system may serve as an
experimental  prototype for those types of problems.

Our discussion here has concentrated on  deriving and analyzing
convenient partition function expressions that one
can also use in quantum Monte Carlo simulations.
We have done some work in this direction,  but we must say that the problem 
is still highly
non trivial because of the form of the kernels in the effective action.
However, we expect to further unravel interesting physics for this
problem in future.

\section{\bf Conclusions}
\label{sec:conclusions}
In this paper we have introduced and presented results 
for a model of two capacitively
coupled quantum Josephson junction arrays. 
This is a difficult problem but one that
promises to lead to interesting new physics.
We have first derived 
a semiclassical expression for an effective Hamiltonian, that
allowed us to study the change in the critical temperature for {\it each}
array.  We used two types of variational actions that permitted the 
evaluation of critical temperature shifts as a function of
the inter-layer coupling capacitance. The main qualitative 
result is that an increase in the interaction capacitance 
increases phase coherence in the arrays. Next we considered the
interesting case where one array is quantum phase dominated and the
other Cooper pair charge dominated. 
Here we extended the one-array work of Fazio {\sl et al.} 
\cite{fazio-schon-91} to the capacitively coupled two-array 
 problem.  We wrote an effective action in terms of four interacting 
imaginary time Coulomb-like  gases, and derived an effective Hamiltonian 
for the coupled system. The effective Hamiltonian is
dually symmetric between charge and vorticity in form but with 
complicated kernels. In the simplified case where one array has one 
vortex and the other one  charge, we showed that their interaction has 
a minimal gauge-like coupling. This interaction is, however, nonlocal 
in the gauge field.

Finally, this type of system holds the promise to lead to a variety
of novel experimentally observable macroscopic quantum phenomena. 
In particular, the vortex-charge interaction discussed at the end 
of this paper deals with the interplay of quantum-classical effects, 
and may lead to possible fractional statistics analogies to the 
fractional quantum Hall effect.

\acknowledgments
This work has been partially supported  by $NSF$ grant DMR-9521845.
%

%
%


\end{document}